\documentclass[twocolumn,showpacs,preprintnumbers,amsmath,amssymb,floatfix]{revtex4}
\usepackage{graphicx}
\usepackage{dcolumn}
\usepackage{bm}
\usepackage{color}

\begin{document} 

\title{Optical control of electron spin coherence in CdTe/(Cd,Mg)Te
quantum wells}

\author{E.~A.~Zhukov$^{1,2}$, D.~R.~Yakovlev$^{1,3}$, M.~M.~Glazov$^{3}$,
L.~Fokina$^1$, G.~Karczewski$^4$, T.~Wojtowicz$^4$,
J.~Kossut$^4$, and M. Bayer$^1$} \affiliation{$^1$ Experimentelle
Physik 2, Technische Universit\"at Dortmund, 44221 Dortmund,
Germany} \affiliation{$^2$ Faculty of Physics, M.~V.~Lomonosov
Moscow State University, 119992 Moscow, Russia} \affiliation{$^3$
Ioffe Physical-Technical Institute, Russian Academy of
Sciences, 194021 St.~Petersburg, Russia} \affiliation{$^4$
Institute of Physics, Polish Academy of Sciences, PL-02668 Warsaw,
Poland}
\date{\today}

\begin{abstract}
Optical control of the spin coherence of quantum well electrons by
short laser pulses with circular or linear polarization is studied
experimentally and theoretically. For that purpose the coherent
electron spin dynamics in a \textit{n}-doped CdTe/(Cd,Mg)Te quantum well
structure was measured by time-resolved pump-probe Kerr rotation,
using resonant excitation of the negatively charged exciton (trion)
state. The amplitude and phase shifts of the electron spin beat
signal in an external magnetic field, that are induced by laser
control pulses, depend on the pump-control delay and polarization of
the control relative to the pump pulse. Additive and non-additive
contributions to pump-induced signal due to the control are isolated
experimentally. These contributions can be well described in the
framework of a two-level model for the optical excitation of the resident
electron to the trion.
\end{abstract}


\maketitle
\section{Introduction}
\label{sec:1}

Semiconductor spintronics relies on the possibility to control
electron spins by non-magnetic methods so that high-frequency
manipulation on time scales approaching the pico- and femtosecond
ranges, well below the coherence time, becomes
feasible~\cite{Spinbook,QBbook,Spintronics}. To this end optical
methods have been considered to be most promising. Substantial
experimental and theoretical efforts have been directed towards
studies addressing optical orientation of electron spins as well as
generation and control of electron spin coherence in semiconductor
nanostructures.

Pump-probe techniques are very convenient tools to study coherent
spin dynamics~\cite{Ch6}. Thereby a circularly polarized pump pulse,
typically with duration between 100~fs and 1~ps, generates the
electron spin orientation, which is subsequently monitored by the
weaker linearly polarized probe pulse delayed relative to the pump
pulse. The rotation of the probe polarization plane measured, e.g.,
in transmission (Faraday rotation) or reflection (Kerr rotation)
geometry is directly proportional to the electron spin polarization
along the optical axis. In a perpendicular external magnetic field
the coherent spin precession of the electrons can be monitored
giving access to the electron spin dephasing times.

Possibilities of optical rotation of the electron spin to reach all
points on the Bloch sphere by the spin vector have been widely
discussed. But only very recently this goal has been achieved for a
single quantum dot~\cite{Press08,Ber08}, an ensemble of singly
charged (In,Ga)As quantum dots~\cite{Gre09NP} and a CdTe/(Cd,Mg)Te
quantum well (QW)~\cite{Phelps09}. In these experiments care was
taken that only the spin coherence initiated by the pump is
manipulated, but no additional spin coherence is created by the control.
This was achieved when the control energy was either detuned from the
pump energy, so that the pulses have no spectral overlap, or by means of $2\pi$
control pulses. It was demonstrated that the polarization vector undergoes a
full revolution on the Bloch sphere.

In this paper we report on a different regime, where the control and
pump photon energies coincide. It was shown that in this regime the
spin coherence of Mn spins in CdTe/(Cd,Mn)Te QWs as a result of the
optical excitation shows additive contributions of the pump and
control pulses~\cite{Akimoto98}. It can be enhanced or suppressed by
proper choice of the control polarization and time delay relative to
the pump pulse. Here we investigate the electron spin coherence in
CdTe/(Cd,Mg)Te QWs containing a low density electron gas, for which
spin coherence is generated by resonant excitation of the negatively
charged exciton (trion) resonance~\cite{Ast05,Zhukov07}. We found
that the control effect is determined by additive and non-additive
mechanisms, whose relative strengths depend on the electron spin
polarization initiated by the pump. Surprisingly, a linearly
polarized control pulse causes a very efficient suppression of the
electron spin coherence, while excitation with such a pulse does not
lead to any spin polarization. The developed quantitative theory
allows us to explain these experimental data quantitatively.

The paper is organized as follows. After introducing the experiment
in Section~\ref{sec:2}, in Section~\ref{sec:3} experimental results
for the optical control with circularly and linearly polarized
control pulses are described. Also a qualitative model of the effect
of a linearly polarized control on the signal suppression is
presented. Section~\ref{sec:9} is devoted to quantitative
theoretical considerations based on a two-level system for the
electron-trion optical excitation. The experimental results are
compared with the modeling in Sec.~\ref{sec:disc}. Here we also
discuss possible reasons for deviations between the experiment and
theory at high control powers.

\section{Experimental techniques}
\label{sec:2}

The studied CdTe/Cd$_{0.78}$Mg$_{0.22}$Te QW heterostructure (sample
031901D) was grown by molecular-beam epitaxy on top of a 2 $\mu$m
CdTe buffer layer deposited on a $(100)$-oriented GaAs substrate. It
contains 5 periods, each of them consisting of a 110-nm-thick
Cd$_{0.78}$Mg$_{0.22}$Te barrier and a 20-nm-thick CdTe QW. An
additional 110-nm-thick barrier was grown on top of this layer
sequence to reduce the influence of surface charges on the confined
electronic states in the QWs. The barriers include 15~nm~layers
doped by Iodine donors, which are separated by 20~nm~spacers from
the QWs. These modulation doped layers provide electrons
being collected in the QWs, where two-dimensional electros gases (2DEGs) with a low density of about
$n_e= 2\times 10^{10}$~cm$^{-2}$ form. This sample has a slightly
larger electron density compared to its partner sample 031901C
($n_e= 1.1\times 10^{10}$~cm$^{-2}$) grown on the same substrate by
a wedge growth technique~\cite{Woj98} which has been studied in
Ref.~\cite{Zhukov07}. The optical properties of both samples are,
however, similar to each other, see Refs.~[\onlinecite{Zhu06a,
Yak07, Zhukov07}] for details.

The measurements were performed in magnetic fields up to 7~T applied
perpendicular to the structure growth axis,
$\mathbf{B}\perp\mathbf{z}$ (Voigt geometry). The sample was
immersed in pumped liquid Helium at a temperature of $T=1.9$~K.

Time-resolved pump-probe Kerr rotation (KR) technique was used  to
study the coherent spin dynamics of the resident QW
electrons~\cite{Zhukov07}. Two mode-locked Ti:Sapphire lasers
synchronized with each other generated the 1.5~ps pump and control
pulses (spectral width of about 1~meV) at a repetition frequency of
75.6~MHz. The probe beam was split off from the pump laser, as
sketched in Fig.~\ref{fig:1}(a). For the experiments reported here
both lasers were tuned to the same photon energy corresponding to
the trion resonance.

The electron spin coherence was excited by the pump and control
pulses, for which different polarization configurations were used:
The control was either co- or cross-circularly polarized with
respect to the pump of fixed circular $\sigma^+$ polarization, or it
was linearly polarized. The induced spin coherences were monitored by the
reflected linearly polarized probe pulse, for which the angle of
Kerr rotation was measured by a balanced photodetector interfaced by
a lock-in amplifier, after sending it through a polarization
sensitive Glan-Thompson beam splitter. The time delay between pump
and probe pulses could be varied up to 7~ns by a mechanical delay
line. A second delay line was used to set a fixed delay of the
control pulse relative to the pump pulse. This delay could be
changed up to $ t_{pc}\leq 2$~ns in order to tune the phases of the
spin coherences initiated pump and control with respect to each
other.

Two protocols of pump and control beam modulation were used. First
we present experiments, where the signals are mainly given by the
additive effect of the pump and control actions. Here both pump beam
and control beam were modulated by a chopper at a frequency of
1~kHz, so that the detected Kerr rotation signal reflects the effect
of both beams.
These measurements are described in Secs.~\ref{sec:4}-\ref{sec:6}.

In order to study the "non-additive" effect of the control on the
pump induced signal we used a protocol in which only pump beam was
modulated. It was sent through a photoelastic modulator operated at
50~kHz frequency so that the polarization was modulated between
$\sigma^+$ and $\sigma^-$. The polarization of the control beam was
constant in time.
The Kerr rotation signal was detected at the pump modulation
frequency of 50~kHz, which allows us to suppress the additive
contribution to the electron spin polarization induced by the
non-modulated control beam. These results are reported in
Sec.~\ref{sec:8}.

\section{Experimental results and discussion}
\label{sec:3}

\begin{figure}[hbt]
\includegraphics*[width=8 cm]{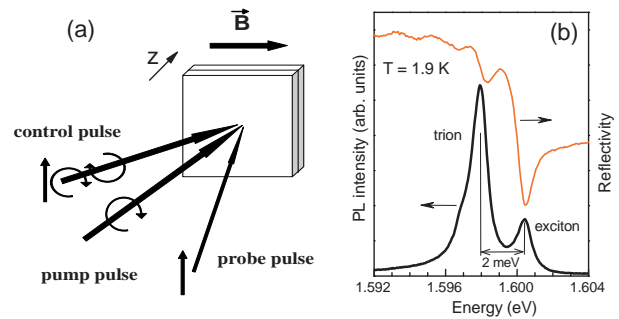}
\caption[] {(Color online) (a) Scheme of the three-pulse
time-resolved Kerr rotation experiment. (b) Photoluminescence and
reflectivity spectra of a 20-nm-thick CdTe/Cd$_{0.78}$Mg$_{0.22}$Te
QW. PL was measured under nonresonant \emph{cw} excitation with
photon energy of 2.33~eV.} \label{fig:1}
\end{figure}

Photoluminescence (PL) and reflectivity spectra of the studied QW
structure are shown in Fig.~\ref{fig:1}(b). The heavy-hole exciton
(X) and negatively charged trion (T) resonances are clearly seen as
minima in the reflectivity spectrum and as lines in the PL spectrum.
They are separated by 2~meV, which corresponds to the trion binding
energy~\cite{Ast05,Zhukov09}. The broadening of these lines is
mainly due to exciton and trion localization on QW width
fluctuations. From the relative oscillator strengths of the exciton
and trion resonances in the reflectivity spectrum we evaluate the
resident electron concentration in the QW as  $n_e= 2\times
10^{10}$~cm$^{-2}$ using the method described in Ref.~\cite{Ast02a}.

\begin{figure}[hbt]
\includegraphics*[width=7 cm]{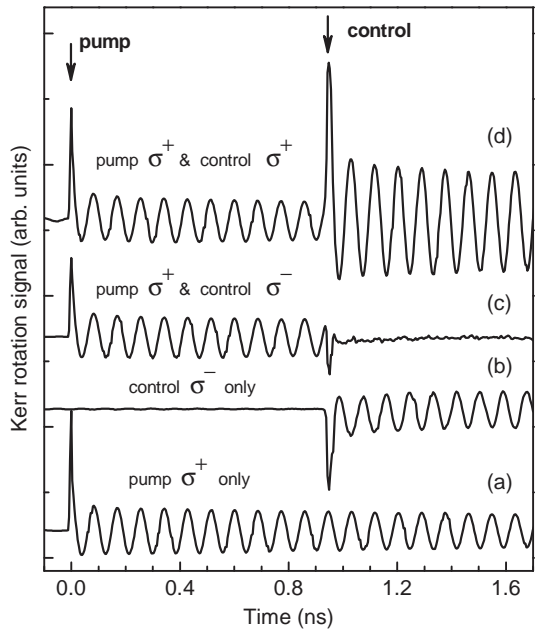}
\caption[] {Kerr rotation signals measured by degenerate
pump-control-probe spectroscopy resonant with the trion energy:  (a)
Only pump pulse with $\sigma^+$ polarization and density of 0.3
W/cm$^2$. Evaluated $T_2^*=4.2$~ns. (b) Only control pulse with
$\sigma^-$ polarization and density of 0.3 W/cm$^2$. (c) $\sigma^+$
pump pulse and $\sigma^-$ control pulse joint excitation. (d)
$\sigma^+$ pump pulse and $\sigma^+$ control pulse joint excitation.
$B=0.5$~T, $T=1.9$~K. For (b), (c) and (d) $t_{pc}= 0.96$~ns and
$\varphi = 0$.} \label{fig:2}
\end{figure}

A typical Kerr rotation signal measured at a magnetic field of 0.5~T
is shown in Fig.~\ref{fig:2} by curve (a). The $\sigma^+$ circularly
polarized pump pulse hits the sample at zero time delay and induces
coherent spin precession of the resident electrons about the
external magnetic field. The precession is reflected by the
periodically oscillating Kerr signal amplitude $K(t)$. The
oscillation period corresponds to the electron Larmor frequency
$\omega_e = \mu_B g_e B / \hbar$ with an electron g-factor $|g_e| = 1.64$, 
which is in
good agreement with literature data~\cite{Sirenko}. Here $\mu_B$ is the Bohr magneton. The g-factor value was
obtained from fitting the experimental data by an exponentially
decaying harmonic function~\cite{Ch6}
\begin{equation}
\label{Kerr_signal}
 K(t) = A\ \exp\left(-\frac{t}{T^*_2}\right) \cos(\omega_e t).
\end{equation}
Here $A$ corresponds to the signal amplitude, $T^*_2$ is the
dephasing time describing the signal decay. The evaluated dephasing
time $T_2^*=4.2$~ns is considerably longer than the trion
recombination times in the range of 30-100~ps in CdTe-based
QWs~\cite{Zhukov07}, which allows us to ascribe the Kerr signal to
resident electrons.

\subsection{Effect of circularly polarized control on signal amplitude}
\label{sec:4}

We turn now to the main topic of the present paper, namely the
effect of control pulses, delayed by a time $t_{pc}$ relative to the
pump pulse, on the electron spin coherence generated by the pump.
The modifications induced by the control depend critically on the
reduced phase $\varphi$ with which the control hits the pump excited
electron spin coherence. This reduced phase is defined as $\omega_e
t_{pc} = \varphi + 2\pi N$, where $N$ is an integer corresponding to
the number of full spin precession periods during the pump-control
delay, and $0 \leqslant \varphi < 2\pi$.

We describe first the pump-probe experiments in which a circularly
polarized control was used. We also focus on the signal amplitude
modifications induced by the control, the changes of the phase are
discussed in Sec.~\ref{sec:5}. To that end we adjust the delay
$t_{pc}$ such that phase $\varphi = 0$ is achieved, when the Kerr
signal amplitude $K(t)$ is maximum. At a magnetic field of 0.5~T
this condition is fulfilled e.g. at $t_{pc}=0.87$~ns or $t_{pc}=0.96$~ns. 
The latter example one
can see in Fig.~\ref{fig:2} by comparing curves (a) and (b). For
co-polarized pump and control pulses (both $\sigma^+$) of the same
power the Kerr signal is enhanced about twice after control action,
see curve (d). This is the expected result, as in this case the
electron spin polarization generated by the control has the same
orientation as the one generated by the pump after a few full
revolutions about the field. In contrast, cross-polarization of the
pump ($\sigma^+$) and control ($\sigma^-$) pulses leads to full
suppression of the electron spin precession signal, as shown by
curve (c). In this case the electron polarizations generated by the
pump and control are antiparallel and compensate each other.

Note, that for the low excitation density regime presented in
Fig.~\ref{fig:2} only a small fraction of the resident electrons is
affected by the pump and control pulses. In this case the joint
action of the pump and control can be described such that each of
them generates spin coherence for two independent subensembles of
electrons. The experimentally measured Kerr rotation signal results
from their independent contributions, which make either additive
or subtractive effect on the observed signal. Note, that a very similar
behavior has been previously reported for the Mn spin coherence in
CdTe/(Cd,Mn)Te QWs~\cite{Akimoto98}.

\begin{figure}[hbt]
\includegraphics*[width=8 cm]{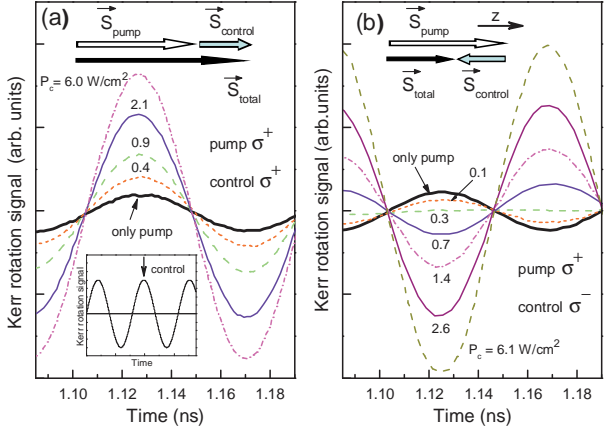}
\caption[] {(Color online) Kerr rotation signals measured for
various control pulse power ($P_c$) at $B = 0.5$~T. Pump is
$\sigma^+$ polarized with power of $P_p = 0.3$~W/cm$^2$: (a)
$\sigma^+$ control pulse; (b) $\sigma^-$ control pulse. Arrow in the
inset marks time moment of control pulse arrival at $t_{pc}=
0.87$~ns, which corresponds to $\varphi = 0$. Arrows in the panels
show schematically the contributions to electron spin polarization
induced by the pump ($\textbf{S}_{pump}$), the control
($\textbf{S}_{control}$) and the result of their joint action
($\textbf{S}_{total}$).} \label{fig:3}
\end{figure}

Detailed results for the effect of control power on the Kerr signal
amplitude for co- and cross-polarizations of pump and control are
given in Fig.~\ref{fig:3}. The phase for control pulse arrival was
chosen to be $\varphi = 0$, as in Fig.~\ref{fig:2}. Therefore, the
spin polarizations induced by the pump ($\textbf{S}_{pump}$) and the
control ($\textbf{S}_{control}$) are either parallel or antiparallel
to each other for co- and cross-polarizations, respectively. The
resultant polarization ($\textbf{S}_{total}$) along the $z$-axis is
reduced or increased, as shown schematically in the corresponding
panels of Fig.~\ref{fig:3}.

\begin{figure}[hbt]
\includegraphics*[width=7 cm]{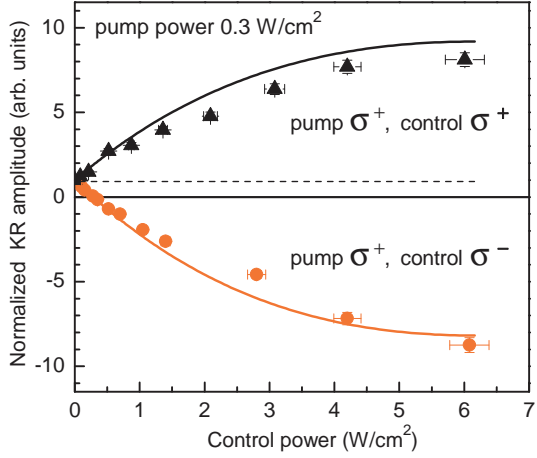}
\caption[] {(Color online) Amplitude of Kerr rotation signal as
function of control power. The amplitude is taken after the time
moment of control pulse arrival for co- and cross-polarization of
pump and control: triangles - pump $\sigma^+$ and control $\sigma^+$
(regime of amplification); circles - pump $\sigma^+$ and control
$\sigma^-$ (regime of suppression). $B = 0.5$~T, $t_{pc}=0.87$~ns,
and $\varphi = 0$. Solid lines show calculated amplitudes of the
signals according to Eq.~\eqref{circ_after}, with details presented
in Sec.~\ref{sec:9} and discussed in Sec.~\ref{subsec:comp}.}
\label{fig:7}
\end{figure}

The Kerr amplitude increases for the co-polarized configuration
shown in Fig.~\ref{fig:3}~(a), in line with the intuitive expectations. It
decreases for the cross-polarized case given in Fig.~\ref{fig:3}~(b), crosses
the zero level when the control power becomes about equal to the
pump power and then shows increasing negative values. These
dependencies can be seen in detail in Fig.~\ref{fig:7}, where the
dependence of the Kerr amplitude on control power is plotted. To
determine the spin beat amplitudes the signals after the control
pulse arrival were fitted by Eq.~\eqref{Kerr_signal}. Triangles and
circles give the experimental data for co- and cross-polarized pump
and control pulses, respectively. The absolute changes of the KR
amplitudes relative to the dashed line are larger for the
cross-polarized configuration. This results from the faster
saturation of the electron spin for co-polarized excitation compared
to the cross-polarized case. We will discuss that in more details in
Sec.~\ref{subsec:comp}.

\subsection{Effect of circularly polarized control on signal phase}
\label{sec:5}

\begin{figure}[hbt]
\includegraphics*[width=7.5 cm]{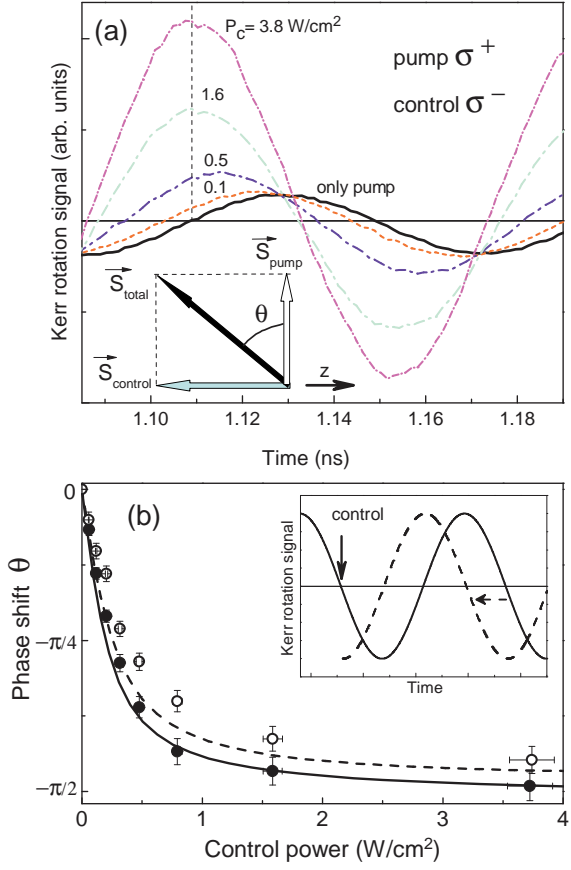}
\caption[] {(Color online) (a) Kerr rotation signals measured for
different control powers ($P_c$): pump ($\sigma^+$, 0.15~W/cm$^2$),
control ($\sigma^-$). Insert shows scheme of the excited electron
spin polarizations. $B = 0.5$~T, $t_{pc}=0.89$~ns, and $\varphi =
\pi / 2$. (b) Phase shift of Kerr rotation signal as function of
control power. Closed circles show phase shift determined from
fitting the experimental data using Eq.~\eqref{Kerr_signal_phase}.
Open circles show values calculated from the experimental data using
the spin composition model Eq.~\eqref{phase_add}. Lines show the
phase of spin beats calculated from the microscopic model,
Eq.~\eqref{theta:all} (solid line) and from the additive model
(dashed lines) Eq.~\eqref{theta:add}, for details refer to
Sec.~\ref{subsec:circ}.  Inset shows schematically the modification
of the pump-induced signal (solid line) by the control pulse
arriving such that $\varphi = \pi / 2$, inducing a phase shift of
the resultant signal (dashed line).} \label{fig:4}
\end{figure}

When the control pulse acts on the pump induced polarization at an
arbitrary phase $\varphi$, not only the amplitude of the Kerr
rotation signal changes, but also the phase will be shifted by an
angle $\theta$ after the control pulse arrival. Corresponding
experimental data are shown in Fig.~\ref{fig:4}(a), where we chose
cross-polarization for pump and control and $\varphi =  \pi / 2$.
The insert in Fig.~\ref{fig:4}(b) shows schematically that for these
experimental conditions the signal after the control pulse is
expected to show a negative phase shift, i.e. to shift to earlier
delays. The signal after control pulse arrival can be described by
Eq.~\eqref{Kerr_signal} when replacing $\cos{(\omega_e t)}$ by
$\cos{(\omega_e t+\theta)}$:
\begin{equation}
\label{Kerr_signal_phase}
K(t) = A\ {\exp\left(-\frac{t}{T^*_2}\right)}{\cos}(\omega_e t + \theta).
\end{equation}

In agreement with our qualitative expectations, the signal phase
shown in Fig.~\ref{fig:4}(b) by the filled circles decreases and
saturates at $\theta = -\pi/2$ for control powers strongly exceeding
the pump power.

The open circles in Fig.~\ref{fig:4}(b) show the signal phase
evaluated from the experimental signal amplitudes without and with
control using the simple additive model depicted in the insert of
Fig.~\ref{fig:4}(a). As one can see from scheme the phase shift
$\theta$ is determined in this case of perpendicular orientation of
$\textbf{S}_{pump}$ and $\textbf{S}_{control}$ by
\begin{equation}
\label{phase_add} \theta=\arctan{(S_{control}/S_{pump})}.
\end{equation}
The overall tendency of the dependences shown by the closed and
open circles is the same. However, they deviate considerably from each
other for control powers exceeding $0.5$~W$/$cm$^2$. This evidences
some non-additive contribution of the control to the spin coherence
generated by the pump which we will discuss in detail below.

\begin{figure}[hbt]
\includegraphics*[width=7.5 cm]{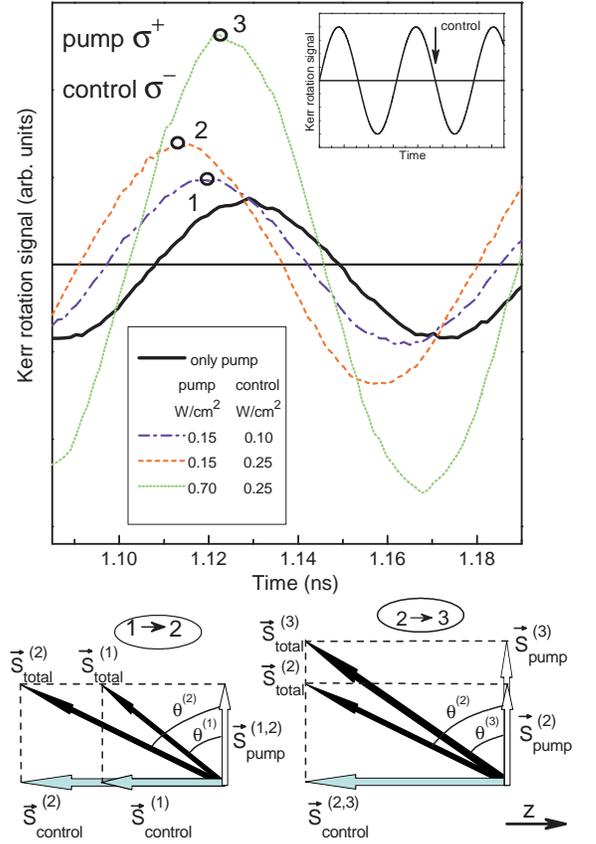}
\caption[] {(Color online) Kerr rotation signals measured at
different pump ($\sigma^+$) and control ($\sigma^-$) powers. $B =
0.5$~T, $t_{pc}=0.89$~ns, and $\varphi = \pi / 2$. Inset illustrates
the control pulse arrival time (arrow) relative to the KR signal
generated by the pump. Two bottom diagrams illustrate the changes of
the phase shift $\theta$ of the KR signal.} \label{fig:5}
\end{figure}

The results in Fig.~\ref{fig:5} have been collected to confirm the
conclusion drawn from the data in Fig.~\ref{fig:4}, that the phase
shift of the Kerr rotation signal is mainly controlled by the ratio
of the pump and control generated spin polarizations,
$S_{control}/S_{pump}$. An increase of the control power for
constant pump power causes a shift of the signal to earlier times,
compare curves 1 and 2. This corresponds to an increase of the phase
shift value, as shown by the left diagram. In turn, a pump power
increase for constant control power (curves 2 and 3 and the right
diagram) induces a signal shift to later times. For the chosen power
densities these transformations are dominated by the additive
mechanism.

\subsection{Effect of linearly polarized control}
\label{sec:6}

\begin{figure}[hbt]
\includegraphics*[width=7 cm]{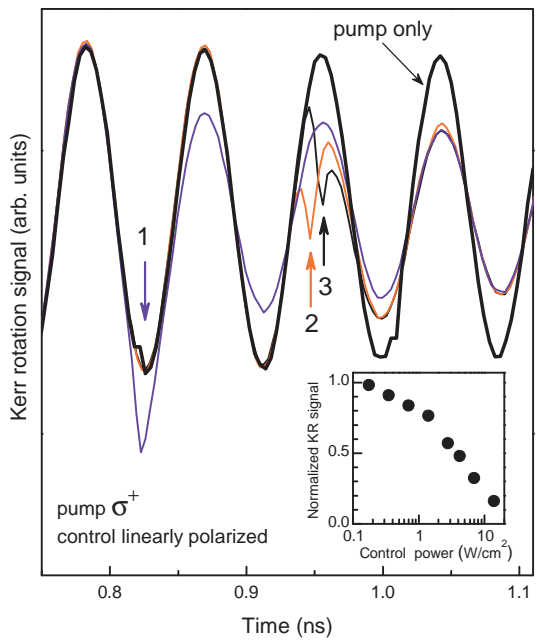}
\caption[] {(Color online) Kerr rotation signals measured at
different time moments of control pulse arrival (indicated by
arrows) for $\sigma^+$ polarized pump with $P_p=2.2$~W/cm$^2$ and
linearly polarized control with $P_c=2.2$~W/cm$^2$. $B = 0.5$~T.
Arrival times of the control pulses are shown by arrows: (1)
$t_{pc}=0.82$~ns, $\varphi = \pi$; (2) $t_{pc}=0.95$~ns, $\varphi = 1.8\pi$; and (3) $t_{pc}=0.96$~ns, $\varphi = 0$.
Insert illustrates suppression of KR signal amplitude with
increasing control power. The amplitude is normalized to its value
without control.} \label{fig:6}
\end{figure}

In our experimental geometry it is not expected that linearly
polarized light would induce any spin polarization of the resident
electrons. Indeed, we did not find any signal for a linearly
polarized pump. However, we observed that the electron spin
polarization induced by a circularly polarized pump is strongly
sensitive to a linearly polarized control. One can see in
Fig.~\ref{fig:6} that irrespective of the delay $t_{pc}$ the Kerr
rotation signal is suppressed by a linearly polarized control. The
suppression effect increases for higher control powers as shown in
the insert. One should note that this effect changes only the signal
amplitude but does not induce any phase shift $\theta$,
independently of $t_{pc}$. The suppression is clearly a non-additive
effect: generation of spin coherence by the control pulse is absent,
but the signal is still modified. These, at first glance, surprising
experimental findings can be explained by the qualitative model
presented in the following section.

\subsection{Qualitative model consideration of linearly polarized control action}
\label{sec:7}



In order to develop a qualitative picture of the spin depolarization
by the \emph{linearly} polarized control we consider the simple
model of a spin ensemble described in Ref.~\cite{Zhukov07}. We
represent the linearly polarized pulse as a superposition of two
circularly polarized ones and assume that at the hit time of the
control pulse there are $n_+$ electrons with spin $z$-component
$1/2$ and $n_-$ electrons with spin $z$-component $-1/2$. We assume
that the control pulse arrives at the maximum ($\varphi=0$) or the
minimum ($\varphi=\pi$) of the pump-induced spin beats, i.e. there
are no in-plane spin components at the moment of control pulse
arrival.

The absorption of the $\sigma^+$ component of the linearly polarized
light generates $n_+ W$ singlet trions by exciting the same number
of $s_z=+1/2$ resident electrons. Here $W$ is the probability of
singlet trion formation per electron due to control pulse action.
Analogously, the $\sigma^-$ component of the linearly polarized
light generates $n_- W$ singlet trions by exciting the same number
of $s_z=-1/2$ electrons. Provided the hole spin-flip time is much
shorter than the trion radiative lifetime the electrons bound to
trions are left unpolarized after trion recombination. Therefore,
the total spin of the ensemble is decreased by
\begin{equation}
\delta S_z  =  S_z^{(a)}-S_z^{(b)} = -\frac{n_+-n_-}{2}W =
-S_z^{(b)} W.
\end{equation}
Here the superscripts $(a)$ and $(b)$ correspond to the spin $z$ component after and before the control pulse arrival, respectively.
The $z$ projection of the total spin of the electron ensemble after
control pulse arrival is given by
\begin{equation}
\label{after:before:simple} S_z^{(a)} = (1-W)S_z^{(b)}.
\end{equation}
Clearly, the probability of singlet trion formation is $0\leqslant W
\leqslant 1$ so that the electron spin after the control pulse is
smaller than the spin before the pulse. It follows therefore that
the linearly polarized pump acts as a depolarizer.

\subsection{Non-additive contribution of control}
\label{sec:8}

In this section we address experimentally the question whether a
circularly polarized control, similar to a linearly polarized one,
can serve as a depolarizer of the induced spin coherence. This will
also allow us to obtain in-depth insight into the non-additive
contribution noted in Sec.~\ref{sec:5}. Our goal here is to study
modifications of the pump-induced spin coherence by the control. For
that one should exclude Kerr rotation signal that is directly caused
by generation of electron spin polarization by the circularly
polarized control. It is possible to suppress this signal by
implementing the second measurement protocol described in
Sec.~\ref{sec:2}. Only the pump beam is modulated in this case and
lock-in detection allows us to exclude the direct contribution of
the unmodulated control to the detected spin polarization.

\begin{figure}[hbt]
\includegraphics*[width=7 cm]{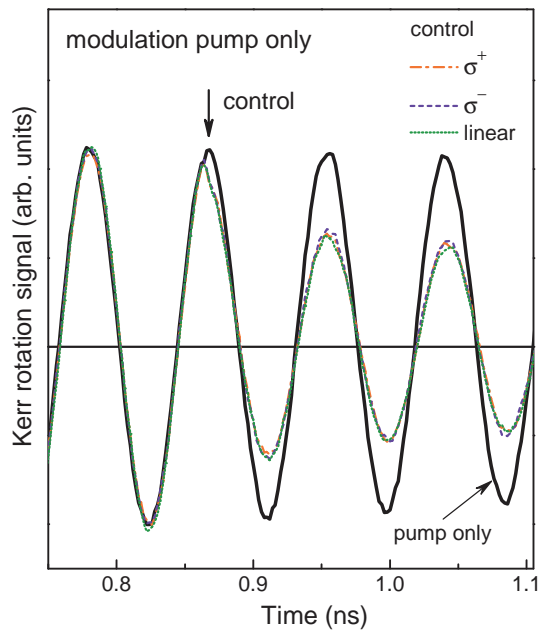}
\caption[] {(Color online) Non-additive effect on Kerr rotation
signals measured for different control polarizations ($\sigma^+$,
$P_c=3.5$~W/cm$^2$) when only the pump beam is modulated
($P_p=0.25$~W/cm$^2$). Arrow indicates the time moment of control
pulse arrival at $t_{pc}=0.87$~ns, $\varphi = 0$. $B = 0.5$~T.}
\label{fig:8}
\end{figure}

One can see in Fig.~\ref{fig:8} that also a circularly polarized
control decreases the Kerr rotation amplitude, similar to the case
of a linearly polarized control. The magnitude of this effect is
identical for $\sigma^+$ and $\sigma^-$ polarization of the control
and is also independent of the control delay $t_{pc}$ (not shown).
It is interesting that the suppression efficiency of the circularly
polarized control is equal to the one for a linearly polarized
control of the same intensity. This suggests that the responsible
mechanism is the same, which is confirmed by the quantitative
analysis given below.

\begin{figure}[hbt]
\includegraphics*[width=7 cm]{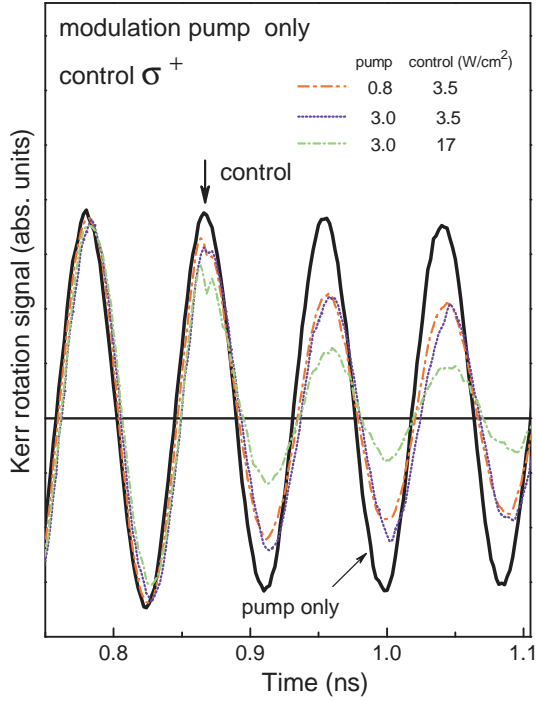}
\caption[] {(Color online) Non-additive effect on Kerr rotation
signals measured at various pump and control ($\sigma^+$) powers
(pump beam modulation only). Signals are normalized by the pump
power to simplify comparison with each other. Thick solid line is
pump only signal for $P_p=0.25$~W/cm$^2$. $B = 0.5$~T,
$t_{pc}=0.87$~ns, and $\varphi = 0$.} \label{fig:9}
\end{figure}

In Figure~\ref{fig:9} the effect of the non-additive contribution is
presented for various pump and control powers. The Kerr rotation
signals are normalized to their maximum amplitudes before control
pulse arrival.  Two conclusions follow from these experimental data.
First, the suppression efficiency increases with increase of the
control power. Second, the suppression efficiency is determined by
the control power only, compare the signal amplitudes for different
pump powers before and after control arrival for the same control
power of 3.5~W/cm$^2$.

\section{Quantitative theory}
\label{sec:9}

The quantitative theory of spin manipulation by a control pulse is
developed following the methods described in Ref.~\cite{Yugova09}.
The electric field of the control pulse can be written as
\begin{equation}
\label{control}
 \bm E(\bm r, t) = E_{\sigma^+}(\bm r,t) \bm o_+  + E_{\sigma^-}(\bm r,t)\bm o_- + {\rm c.c.}\:,
\end{equation}
where $\bm o_\pm$ are the circularly polarized unit vectors
related to the unit vectors ${\bm o}_x \parallel x$ and ${\bm o}_y
\parallel y$ by $\bm o_\pm = (\bm o_x \pm \mathrm i \bm
o_y)/\sqrt{2}$. Here the components $E_{\sigma ^+}$ and
$E_{\sigma^-}$ are proportional to the product of the exponential
function $ \mathrm{exp}(-\mathrm i \omega_{ \mbox{}_{\rm C}} t)$
with $\omega_{\mbox{}_{\rm C}}$ being the control pulse optical
frequency {and a smooth envelope}.

The incident electromagnetic field induces optical transitions
between the electron state and the trion state, creating a coherent
superposition of them. In accordance with the selection rules
$\sigma^+$ circularly polarized light creates a superposition of the
$+1/2$ electron and $+3/2$ trion states, while $\sigma^-$ polarized
light creates a superposition of the $-1/2$ electron and $-3/2$
trion states. In order to describe these superpositions it is
convenient to introduce a four component wavefunction
\begin{equation}
 \label{wave}
\Psi = \left( \psi_{1/2}, \psi_{-1/2}, \psi_{3/2}, \psi_{-3/2}
\right)\:,
\end{equation}
where the $\pm 1/2$ subscripts denote the electron spin projection
and $\pm 3/2$ refer to the spin projection of the hole in the trion.
The electron spin polarization is expressed in terms of $\psi_{\pm
1/2}$ as follows
\begin{eqnarray}
\label{eq:12}
S_z&=&\left(|\psi_{1/2}|^2-|\psi_{-1/2}|^2\right)/2\:,\nonumber\\
S_x&=&\Re(\psi_{1/2}\psi_{-1/2}^*)\:,\nonumber\\
S_y&=&-\Im(\psi_{1/2}\psi_{-1/2}^*)\:.
\end{eqnarray}
Here $\Re$ and $\Im$ are real and imaginary parts, respectively. All excited states of the system, such as e.g. triplet trion states
are neglected. In this respect the model is directly applicable to
the case of a resident carrier strongly localized in a quantum dot
or quantum well imperfection. The role of excited states will be
discussed below, in Sec.~\ref{subsec:circ_strong}.

Further, we assume that the delay between the pump and control
pulses exceeds by far the radiative lifetime of the trion, hence,
just before the control pulse arrival there is a resident electron
with precessing spin but no trion. The state of the system just
before the control pulse arrival corresponds to the non-zero
components $|\psi_{+1/2}|^2 + |\psi_{-1/2}|^2 = 1$ and $\psi_{\pm
3/2}=0$.

Following the method in Ref.~\cite{Yugova09} and introducing smooth
envelopes for the $\sigma^+$ and $\sigma^-$ polarized components of
the control pulse by
\[
f_\pm(t) = -\frac{\mathrm e^{\mathrm i \omega_{\mbox{}_{\rm C}}
t}}{\hbar}\int \mathsf d(\bm r) E_{\sigma_\pm}(\bm r,t)\mathrm d^3
r\:,
\]
where $\mathsf d(\bm r)$ is the effective transition dipole, see Eq.
(12) in Ref.~\cite{Yugova09}, one may reduce the Schroedinger
equation for the four-component wave function to two independent
differential equations for $\psi_{\pm 1/2}(t)$ which take the
following simple form
\begin{equation} \label{single}
\ddot{\psi}_{\pm 1/2} - \left( \mathrm i \omega'+
\frac{\dot{f}_\pm(t)}{f(t)} \right) \dot{\psi}_{\pm 1/2} +
f^2_\pm(t) \psi_{\pm 1/2} =0\:.
\end{equation}
Here $\omega' = \omega_{\mbox{}_{\rm C}}- \omega_0$ is the detuning
between the control pulse optical frequency and the trion resonance
frequency, $\omega_0$. This simple form of Eqs.~\eqref{single}
follows from (i) disregarding other excited states of the system and
(ii) neglecting the control pulse duration compared to the trion
lifetime and the electron spin precession period in magnetic field.
Below we discuss the cases of linearly and circularly polarized
control pulses.

\subsection{Linearly polarized control}
\label{subsec:lin}

In case of a control pulse linearly polarized along the $x$ axis the
circular components of the pulse envelope function can be written as
\begin{equation}\label{pulse:lin}
 f_\pm(t) =  \frac{\mu}{\sqrt{2}\cosh{\left(\frac{\pi t}{\tau_p}\right)}},
\end{equation}
where the factor $1/\sqrt{2}$ is introduced for convenience, $\mu$
characterizes the amplitude of the control pulse and $\tau_p$ is its
duration. The pulse area is defined as $\Theta=2\mu\tau_p$. The
solution of Eqs.~\eqref{single} can be recast as~\cite{Yugova09}
\begin{eqnarray}\label{after}
\psi_{1/2}(+\infty) = \psi_{1/2}(-\infty) Q_l\mathrm e^{\mathrm i \Phi_l}, \nonumber\\
\quad  \psi_{-1/2}(+\infty) = \psi_{-1/2}(-\infty) Q_l\mathrm
e^{\mathrm i \Phi_l},
\end{eqnarray}
where the constants $Q_l$ and $\Phi_l$ describe the transformation
of the wavefunction under action of the linearly polarized pulse.
For the case of a Rosen\&Zener pulse, Eq.~\eqref{pulse:lin}, one has
\begin{eqnarray}\label{Ql}
 Q_l^2 = 1 - \frac{\sin^2{(\Theta_l/2)}}{\cosh^2{(\pi y)}},
\end{eqnarray}
where $\Theta_l = 2\mu\tau_p/\sqrt{2}$ is the effective area of each
circularly polarized component of the control pulse, and $y =
\omega'\tau_p/(2\pi)$. The expression for the constant $\Phi_l$ is
rather bulky and is therefore not given here, see Eq.~{(26)} in
Ref.~\cite{Yugova09}.

\begin{figure}[hpt]
 \includegraphics*[width=7.5 cm]{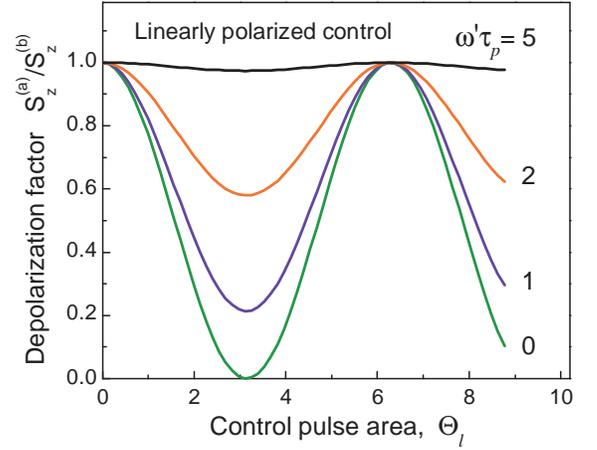}
\caption{(Color online) Depolarization factor $Q^2_l$ as function of
control pulse area $\Theta_l$ calculated for different detunings. }
\label{fig:lin}
\end{figure}

Using the definitions of the spin components, Eqs.~\eqref{eq:12},
one can readily obtain from Eq.~\eqref{after} that the spin vector
of an electron after the control pulse, $\bm S^{(a)}$,  is connected
with the electron spin vector before the control pulse arrival,
 $\bm S^{(b)}$, by
\begin{equation}\label{s_after}
 \bm S^{(a)} = Q_l^2 \bm S^{(b)},
\end{equation}
i.e. the spin vector before the control pulse is simply multiplied
by some nonnegative quantity $Q_l^2 \leqslant 1$. If the electron is
left behind unpolarized after trion decay, i.e. when the trion
lifetime is longer than the hole spin relaxation time, then the
total spin of the electron ensemble is decreased, in agreement with
the simplified Eq.~\eqref{after:before:simple} obtained from
qualitative arguments. The dependence of the depolarization factor
$Q_l^2$ on the control pulse area for different detunings between
the trion resonance and the control optical frequencies is shown in
Fig.~\ref{fig:lin}. The depolarization efficiency shows Rabi oscillations and is larger for
small detunings.

In the case of small control power {effective pulse area} $\Theta_l
\ll 1$, and for negligible detuning between the control pulse and
the trion resonant frequency, $y\ll 1$,  one can represent $Q_l^2$
in Eq.~\eqref{Ql} as
\begin{equation}\label{s_after1}
 Q_l^2  \approx 1 - \frac{(\mu \tau_p)^2}{2}.
\end{equation}

\subsection{Circularly polarized control}\label{subsec:circ}

Now we turn to the case of circularly polarized control pulses. For
a $\sigma^+$ polarized control pulse the envelope function
\begin{equation}\label{pulse:circ}
 f_+(t) =  \frac{\mu}{\cosh{\left(\frac{\pi t}{\tau_p}\right)}},\quad f_-(t)=0.
\end{equation}
The time integrated intensities of the circularly polarized pulse
$\propto \int_{-\infty}^\infty [f_+^2(t) + f_-^2(t)]\mathrm dt$ and
of the linearly polarized pulse, Eq.~\eqref{pulse:lin}, are the
same.

Making use of Ref.~\cite{Yugova09} we obtain the following expressions
which link the spin components before and after control pulse
arrival:
\begin{eqnarray}
 S_z^{(a)} = \mp \frac{1-Q_c^2}{4}  + \frac{Q_c^2+1}{2} S_z^{(b)}, \label{circ_after}\\
S_x^{(a)} =  Q_c\cos{\Phi_c} S_x^{(b)} \pm Q_c\sin{\Phi_c} S_y^{(b)},\\
S_y^{(a)} =  Q_c\cos{\Phi_c} S_y^{(b)} \mp Q_c\sin{\Phi_c} S_x^{(b)}.
\end{eqnarray}
Here the upper signs of $\mp$ and $\pm$ correspond to a $\sigma^+$
polarized control and the lower signs to a $\sigma^-$ polarized
control. The constant $Q_c$ is given by
\[
 Q_c^2 = 1 - \frac{\sin^2{(\Theta_c/2)}}{\cosh^2{(\pi y)}},
\]
where $\Theta_c = \sqrt{2} \Theta_l =2 \mu\tau_p$.  For small pulse
areas $\Theta_c \ll 1$ and $y\ll 1$
\begin{equation}\label{s_after2}
 Q_c^2  \approx 1 - (\mu \tau_p)^2.
\end{equation}

\begin{figure}[hpt]
\includegraphics*[width=7.5 cm]{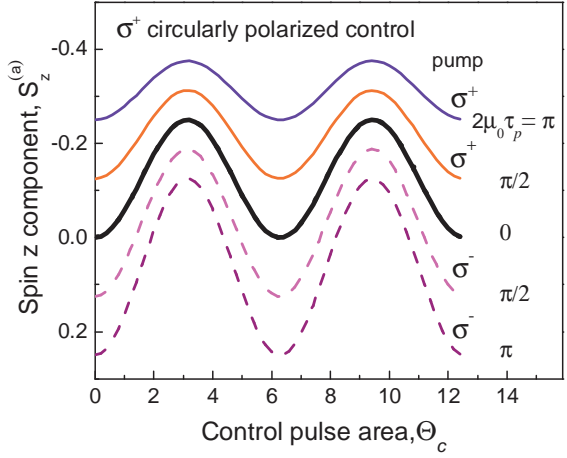}
\caption{(Color online) $S_z^{a}$ component after $\sigma^+$ polarized
control pulse pulse arrival as function of control pulse area
$\Theta_c$.  Different curves correspond to different pump pulse
areas $2\mu_0\tau_p$. Solid curves show the case of co-polarized
pump and control pulses, dashed curves show the case of
cross-polarized pump and control, and the thick solid curve shows the
case of control only ($2\mu_0\tau_p=0$). The control pulse arrives such that
$\varphi=0$. } \label{fig:circ}
\end{figure}

\begin{figure}[hpt]
\includegraphics*[width=7.5cm]{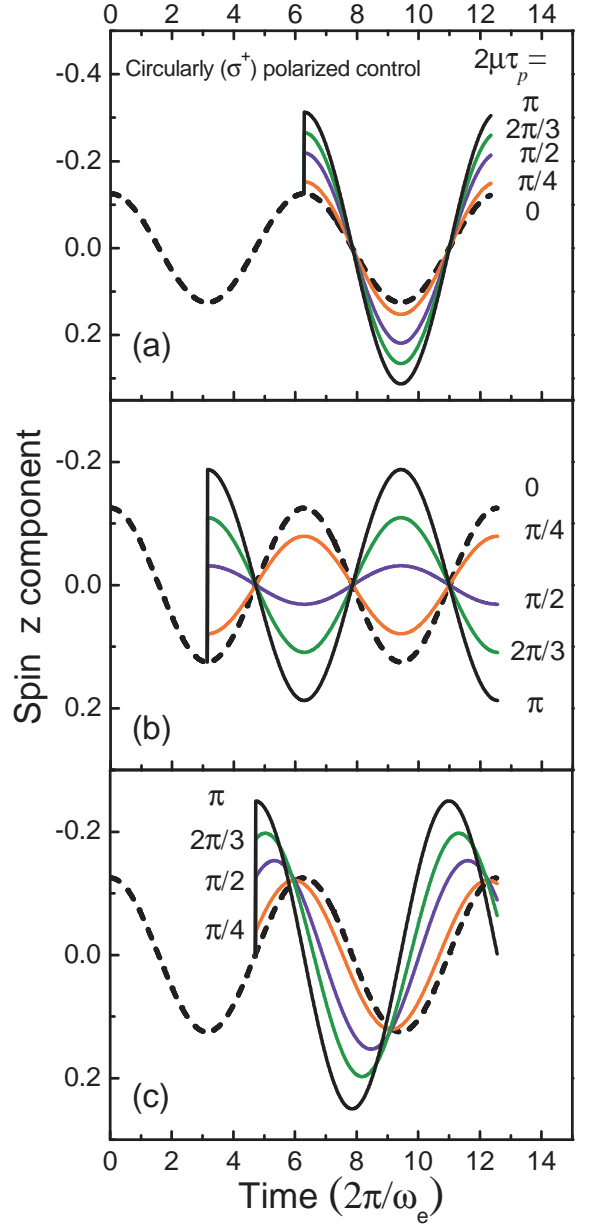}
\caption{(Color online) Time resolved dependencies of $S_z$
calculated for different delays of control pulse arrival (from the
top to the bottom: at beats maximum, $\varphi=0$,  at beats minimum,
$\varphi=\pi$, and at beats zero $\varphi={3}\pi/2$. Different
curves correspond to different amplitudes of the control pulse. The
pump and control pulses are co-circularly polarized ($\sigma^+$).}
\label{fig:phase}
\end{figure}

The modification of the spin $z$ component by a $\sigma^+$ control
pulse for different pump pulse areas are shown in
Fig.~\ref{fig:circ}. Each curve shows the control pulse area
dependence for a fixed pump pulse area $\Theta_0 = 2\mu_0\tau_p$,
where $\mu_0$ is the amplitude of the pump pulse envelope as defined
in Eq.~\eqref{pulse:circ}. Rabi oscillations with period $2\pi$ are
clearly seen. Here we assumed that the control pulse arrives at
$\varphi=0$, i.e. in the same phase as the pump pulse. Note, that a
$\sigma^+$ polarized pump results in an electron spin $z$-projection
$S_z<0$, and corresponds to positive values of the measured Kerr
rotation signal, $K(t)$. For convenient comparison of the
theoretical and experimental results we invert the direction of the
axis of the ``spin $z$ component'' in the theoretical figures. Here
and below the electron spin dephasing is completely neglected in the
calculations. The modification of the spin component $S_z^{(a)}$
comprises both additive and non-additive contributions.
Interestingly, for co-polarized pump and control (solid lines) the
modification is weaker compared with the cross polarized
configuration (dashed lines). This is because the absolute spin
value $|S_z|$ is limited by $1/2$ and when the spin projection is
closer to $-1/2$ (pump and control are co-$\sigma^+$ polarized) the
effect of the control pulse is weaker.

Figure~\ref{fig:phase} shows the time dependencies of the spin $z$
component calculated for different moments of control pulse arrival,
i.e. for different phases of the electron spin precession generated
by the pump: $\varphi=0,\pi$, and $3\pi/2$. The pump and control
pulses are co-circularly polarized. The different curves in each
panel correspond to different areas of the control pulse.

One can see from Eq.~\eqref{circ_after} that there are two
contributions to the spin $z$ component of an electron after
{circularly polarized} control pulse arrival. The first contribution
is an additive one: it changes its sign upon reversal of the
circular polarization of the control pulse and it does not depend on
the spin state before control pulse arrival. For weak control power,
$\mu\tau_p\ll 1$, and negligible detuning, $y\ll 1$, the additive
part to $S_z^{(a)}$ is given by, see Eq.~\eqref{s_after2}
\begin{equation}
\label{Sz_2} \mp \frac{1-Q_c^2}{4} \approx \mp
\frac{(\mu\tau_p)^2}{4}.
\end{equation}
This additive contribution equals exactly the spin $z$ component
created by a pump pulse of the same power.

Another contribution to the electron spin after control pulse action
is a non-additive one. It can be interpreted as a transformation of
the electron spin by the control pulse. This contribution is given
by
\begin{equation}
\label{non-additive} \frac{Q_c^2+1}{2} S_z^{(b)} \approx
\left[1-\frac{(\mu\tau_p)^2}{2} \right]S_z^{(b)},
\end{equation}
where the last approximate equality holds for weak control power and
small detuning. This {non-additive} contribution is independent of
the circular polarization sign and always decreases the $z$
component of electron spin. The comparison of
Eq.~\eqref{non-additive} with Eqs.~\eqref{s_after} and
\eqref{s_after1} shows that for weak control powers the
depolarization of the electron spin $z$ component by circularly and
linearly polarized light is the same.

The in-plane spin components are also affected by the circularly and
linearly polarized control pulses. The absolute value of the
in-plane spin projection $S_\perp = \sqrt{S_x^2+S_y^2}$ is decreased
by the factor $Q_c \approx 1- (\mu\tau_p)^2/2$ (the latter equality
holds for weak control pulses), similar to the case of a linearly
polarized control. In addition, the detuned circularly polarized control
pulse rotates the in-plane spin by the angle $\Phi_c$ around the
$z$-axis.

\begin{figure}[hpt]
 \includegraphics*[width=7.5cm]{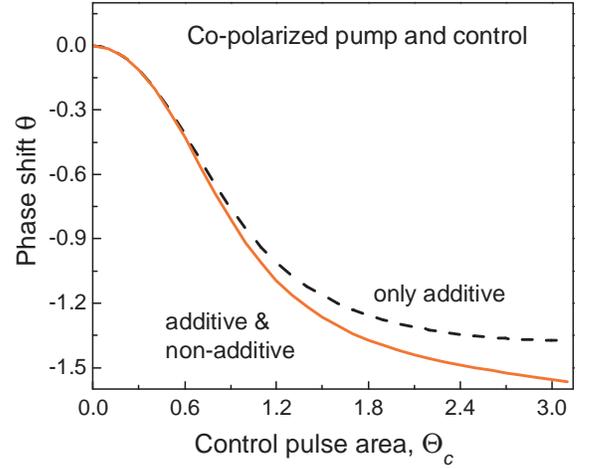}
\caption{(Color online) Phase of the spin beats after control pulse
arrival at zero signal ($\varphi=\pi/2$). Pump and control are
co-circularly polarized. Solid line gives exact calculation, dashed
line is result of an approximate model which accounts for additive
contributions by the control only. The absolute spin value for a
single electron was taken to be $0.05$ which corresponds to a pump
area $\Theta_0=2\mu_0\tau_p=0.93$.} \label{fig:phase0}
\end{figure}

It is noteworthy to analyze the spin beats phase after circularly
polarized control arrival {at $\varphi = \pi/2$ where the signal
amplitude is zero}. In order to calculate the spin beats phase we
assume that the magnetic field is applied along the $x$ axis. We
neglect the detuning between the control pulse optical frequency and
the trion resonance frequency. Hence, the phase shift of the spin
beats induced by the control is given by
\begin{equation}
\label{theta:all}
\theta=\arctan{(S_z^{(a)}/S_y^{(a)})},
\end{equation}
where the spin precession direction was assumed to be clock-wise in
the $(yz)$ plane. The dependence of $\theta$ on the control pulse
area is shown in Fig.~\ref{fig:phase0} by the solid line. We compare
this phase with the results of the simplified additive model, where
we assume that the $y$ spin component is conserved and we take into
account the additive contribution of Eq.~\eqref{circ_after}. The
phase shift in the additive model is
\begin{equation}
\label{theta:add}
\theta'=\arctan{\frac{Q_c^2-1}{4S_y^{(b)}}}.
\end{equation}
This shift is shown by the dashed line in Fig.~\ref{fig:phase0}. The
qualitative behaviors of the two shifts $\theta$ and $\theta'$ are
the same, however, the exact model predicts a stronger phase shift.
This results from the suppression of the in-plane components induced
by the circularly polarized light.

\begin{figure}[hpt]
 \includegraphics*[width=7.5cm]{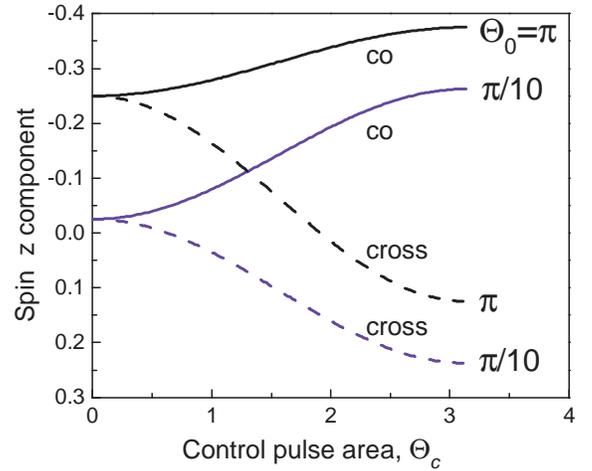}
\caption{(Color online) Electron spin $z$ component  as function of
control pulse area calculated for co- (solid line) and
cross-polarized (dashed line) configurations of pump and control for
two different pump pulse areas, $\Theta_0=\pi/10$ (blue) and
$\Theta_0=\pi$ (black). Phase of control pulse is $\varphi=0$.}
\label{fig:suppr}
\end{figure}

Figure~\ref{fig:suppr} shows the electron  spin $z$ component after
control pulse arrival, calculated as function of control pulse area
for two pump pulse areas and for co- and cross-polarized
configurations. We assumed that the control pulse arrives at phase
$\varphi=0$ of the spin beats. For a weak pump pulse
($\Theta_0=\pi/10$) the additive contribution by the control is
dominant. The modification of the electron spin component is almost
the same in the co- and cross-polarized configurations as it mostly
scales with control power. The maximum absolute value of the
electron spin projection in this case is close to 0.25, in agreement
with Eq.~\eqref{circ_after} for $S_z^{(b)}\ll 1$.

The case of a strong pump pulse, $\Theta_0=\pi$, is different.
Figure~\ref{fig:suppr} shows a strong asymmetry for the induced
polarizations in the co- and cross-polarized configurations. In the
cross-polarized case the change of the spin $z$ component is about
the same as for a weaker pump. In the co-polarized configuration the
control pulse effect is much weaker. This is because the electron
spin coherence generated by the pump pulse is partially suppressed
by the control pulse. For this configuration the maximum absolute
value of the electron spin $z$ component  is $1/4+1/8=0.375$
according to Eq.~\eqref{circ_after}.

\begin{figure}[hpt]
 \includegraphics*[width=7.5cm]{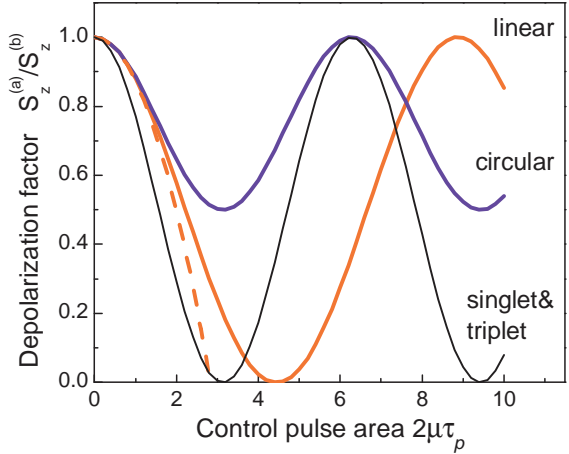}
\caption{(Color online) Suppression of the spin $z$ component by
circularly (blue solid) and linearly (red solid) polarized control
pulses as function of control pulse area. Thin solid line (black)
demonstrates spin suppression by a circularly polarized pulse for
the case when both singlet and triplet trion transitions are excited
with the same probability, Eq.~\eqref{depol:circ:quant} with $\tilde
Q_c = Q_c$. The detuning between the quantum dot trion resonance and
the control optical frequency is zero. Dashed red curve gives the
small amplitude asymptotics.} \label{fig:depol}
\end{figure}

Let us also analyze the non-additive effect by the control pulse for
the case when it arrives exactly in the maximum or minimum of the
spin beats ($\varphi=0$ or $\pi$), i.e. when the in-plane spin
components before control pulse arrival are zero $S_x^{(b)} =
S_y^{(b)}=0$. In this case the electron spin $z$ component is simply
suppressed by the non-additive contribution, in agreement with
Eqs.~\eqref{circ_after} and~\eqref{non-additive}. The efficiency of
the spin depolarization is illustrated in Fig.~\ref{fig:depol}. For
small pulse areas indeed the depolarization is the same for the
linearly and the circularly polarized control. Rabi oscillations are
seen with period $2\pi$ for the circularly polarized control and
with period $2\sqrt{2}{\pi}$ for the linearly polarized control. For
a circularly polarized control the depolarization is weaker and not
complete: one can suppress the spin polarization by no more than a
factor of $2$, while complete depolarization is possible by linearly
polarized light. Note, that complete depolarization is possible for
any arrival phase of the control pulse in case of linear
polarization.

It is worth to mention that the degree of spin suppression by
circularly polarized light is model sensitive. In the following
subsection~\ref{subsec:circ_strong} we demonstrate that the
extension of the model to account for the trion excited states could
result in stronger spin suppression by the circularly polarized
control.

\subsection{Effects of very strong circularly polarized control pulses}
\label{subsec:circ_strong}

Here we analyze briefly the effect of circularly polarized control
pulses of very high intensity. We have seen that the model
description in terms of a two-level model gives a complete
depolarization of the electron spin by linearly polarized light and
partial (by a factor of 2, at most) depolarization by circularly
polarized light. This is because the transition for a given circular
polarization involves just two levels, the ground electron state and
an excited (singlet) trion state. Therefore, only one component of
the electron spin is pumped into the trion state and becomes
subsequently depolarized, while another one is maintained.

There are other possible excited states in the system, e.g. the
triplet trion state, which can be populated by polarized light
absorption. In the classical approach~\cite{Zhukov07} this state can
be considered as an exciton interacting with a resident electron.
Due to the electron spin-flip within a triplet trion a singlet trion
state can be formed.

To analyze the non-additive effect of a circularly polarized control
pulse for the case when the triplet trion/exciton can be
photocreated, we denote the probability of singlet trion formation
via an exciton [as a result of the following process: electron
$-1/2$ + exciton ($-1/2$, $3/2$), afterwards electron spin-flip and
formation of ($-1/2$, $1/2$, $3/2$) or ($-1/2$, $1/2$, $-3/2$)
trion] by $\tilde W$ and the probability of direct singlet trion
formation [$1/2$ electron + photocreated exciton ($-1/2$, $3/2$)
yields ($-1/2$, $1/2$, $3/2$) trion] as $W$.

Let us do the analysis for the experimental scenario of
Sec.~\ref{sec:8} where the pump polarization is assumed to be
modulated while the control is always $\sigma^+$ polarized. If the
electron spin before control arrival is $1/2$ the electron spin
after trion recombination is $(1-W)/2$, because in this case direct
singlet trion formation occurs. If the electron spin after control
arrival is $-1/2$ then its spin after trion recombination is
$-(1-\tilde W)/2$, since formation of a triplet trion/exciton is
required. The detected signal is suppressed compared to the case
without control by the factor
\begin{equation}
\label{depol:circ:class}
S_z^{(a)} = \left(1-\frac{W+\tilde W}{2}\right)S_z^{(b)}.
\end{equation}
At high pump powers both $W$ and $\tilde W$ approach unity (see
Ref.~\cite{Zhukov07}) and the spin after control is completely
erased. Clearly, $W$ approaches $1$ faster since no electron
spin-flip is needed. Therefore one can expect a kind of
``two-stage'' behavior of suppression: first the spin is suppressed
down to the level $(1-\tilde W)/2$ of its value before control pulse
arrival, and further increase of control power yields complete
suppression.

This process can be described quantum mechanically by extending the
wave function $\Psi$, Eq.~\eqref{wave}, to allow for the two triplet
trion states  with total spin projection $\pm 1/2$, formed by two
spin down electrons and a $3/2$ hole or two spin up electrons and a
$-3/2$ hole.
For a $\sigma^+$ control pulse the electron $-1/2$ is excited into a
$1/2$ triplet trion, and, following~\cite{Yugova09} we obtain
\[
\psi_{-1/2}(+\infty) = \tilde{Q}_c \exp{(\mathrm i \tilde{\Phi}_c)} \psi_{-1/2}(-\infty),
\]
\[
\psi_{1/2}(+\infty) = {Q}_c \mathrm \exp{(\mathrm i {\Phi}_c)} \psi_{1/2}(-\infty).
\]
Note that the constants $Q_c$ and $\tilde Q_c$ (as well as $\Phi_c$
and $\tilde{\Phi}_c$) are different because the triplet trion is
usually shifted in energy as compared with the singlet
one~\cite{Ast05}. If we assume, that after trion recombination the
electron is left behind unpolarized, then its spin $z$-component is
given by
\begin{equation}
\label{sz:circl:quant}
S_z^{(a)} = - \frac{\tilde Q_c^2-Q_c^2}{4}  + \frac{\tilde Q_c^2 + Q_c^2}{2} S_z^{(b)},
\end{equation}
Eq.~\eqref{sz:circl:quant} clearly shows that there are both
additive and non-additive contributions to the electron spin $z$
component. The non-additive contribution is
\begin{equation}
\label{depol:circ:quant}
S_z^{(a)} = \frac{\tilde Q_c^2+ Q_c^2}{2}S_z^{(b)}.
\end{equation}
One sees that excitation of the triplet trion state results in
additional suppression of the electron spin polarization.

We note that the probability of singlet trion formation by a short
pulse is $1-Q_c^2$ and the probability of triplet trion formation is
$1-\tilde Q_c^2$. Hence, the quantum and classical approaches are
equivalent to each other if we take $W=1-Q_c^2$ and $\tilde W =
1-\tilde Q_c^2$.

It is instructive to consider two limiting cases:

(i) Only the triplet trion is excited ($\tilde Q_c \ne 0$, $Q_c=0$).
The non-additive spin suppression is fully described by the theory
developed in Secs.~\ref{subsec:lin} and \ref{subsec:circ} by
changing $Q_c \to \tilde{Q_c}$, $\Phi_c \to \tilde \Phi_c$ and
replacing $\mp$ by $\pm$ in Eq.~\eqref{circ_after}. Suppression by
the circularly polarized light is possible by a factor $2$ only,
similar to the situation when only the singlet trion is excited.

(ii) The formation probabilities of the singlet and triplet trions
are the same, $Q_c^2 = \tilde Q_c^2$. The spin suppression factor
for circularly polarized control is given by $Q_c^2$, see
Eq.~\eqref{sz:circl:quant}, i.e. complete depolarization is
possible, see thin solid curve in Fig.~\ref{fig:depol}. It is
remarkable that in this case the depolarization effect by linearly
and circularly polarized controls of the same area are identical.

\section{Discussion}\label{sec:disc}

\subsection{Comparison theory and experiment}\label{subsec:comp}

So far, we have established experimentally and theoretically that
the control pulse has, in general, a two-fold effect on the electron
spin coherence in quantum wells. First, a circularly polarized
control pulse generates additional spins and results in an additive
contribution to the spin beats. Besides, the control pulse effects
the spins that are already polarized by the pump pulse, leading to
suppression of the pump-induced spin coherence. The latter effect is
possible both for circularly and linearly polarized control pulses.

\begin{figure}[hbt]
\includegraphics*[width=8 cm]{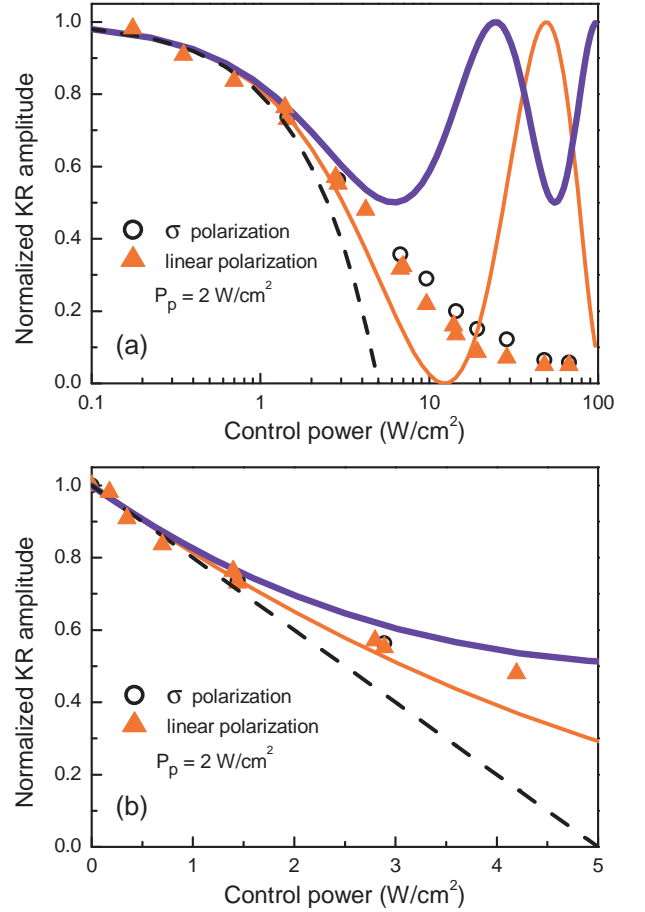}
\caption[] {(Color online) Amplitude of Kerr rotation signal after
control pulse arrival normalized to the amplitude before control
pulse arrival, $S_z^{(a)}/S_z^{(b)}$. Symbols are experimental data
measured for linear polarization of the control pulse (triangles)
and for circular polarization (circles). $B=0.5$~T and $T=1.9$~K.
Curves are theoretical calculations: suppression for linear
polarization (solid red line) and for circular polarization (solid
blue line). Dashed curve shows small power asymptotics.}
\label{fig:ampl}
\end{figure}

To do a quantitative comparison of the experimental and theoretical
results we consider in detail the effect of the spin coherence
suppression by linearly and circularly polarized light.
Figure~\ref{fig:ampl} shows the suppression efficiency, i.e. the
ratio $S_z^{(a)}/S_z^{(b)}$ as function of control pulse power for a
linearly polarized control (closed circles) and a circularly
polarized control (open circles). We focus on the small control
power regime $P_{c} \leqslant 5$ W/cm$^2$ illustrated in detail in
Fig.~\ref{fig:ampl}(b).  In this regime the efficiency of
suppression increases linearly with increasing control pulse power.
Fitting the experimental data by the theoretical model,
Eqs.~\eqref{s_after1} and \eqref{non-additive}, we obtain a relation
between the control pulse power and its area:
\begin{equation}
\label{mutau:p}
P_c = C\Theta_l^2,
\end{equation}
where $C\approx 0.63$~W/cm$^2$ is the fitting parameter. The
theoretical curve corresponding to the limit of $\Theta_l \ll 1$
is shown by the dashed line in Fig.~\ref{fig:ampl}.

The solid thick and thin lines show the suppression efficiency as
function of control power for linear and circular polarizations of
the control pulse, respectively. They are calculated for the whole
range of experimentally used powers by Eqs.~\eqref{s_after1} and
\eqref{non-additive}, using the link between the control power and
its area from Eq.~\eqref{mutau:p}, see Fig.~\ref{fig:ampl}(a). The
theory reproduces the experimental data well for control powers $P
\leqslant 5$ W/cm$^2$. For higher powers the discrepancy between
experiment and theory is large, the reasons for that are discussed
in Sec.~\ref{sec:high}.

It is worth to note that other experimental data recorded at low
pump powers are in good agreement with the theory.
Figure~\ref{fig:7} shows the normalized amplitude of the Kerr
rotation signal measured as function of control power for co- and
cross-polarized configurations. The lines in Fig.~\ref{fig:7} show
the theoretical calculations obtained from Eq.~\eqref{circ_after}
using the relation between the pulse area and control power
Eq.~\eqref{mutau:p} with the same value of $C=0.63$ W/cm$^2$ as in
Fig.~\ref{fig:ampl}. Good agreement between the experimental data
and theoretical curves is seen. Figure~\ref{fig:7} shows that for
co-polarization the amplitude of the signal saturates faster than
for cross-polarization. This is reasonable, because the spin
projection of a single electron is limited by $1/2$. Therefore, in
co-polarization the spin should saturate faster because spin with
projection of the same sign is added and, therefore, the spin
reaches the maximum value faster.

We also address the phase shift of the spin beats $\theta$ as
function of control power, Fig.~\ref{fig:4}(b). The black circles
shows the phases of the Kerr signal after control pulse arrival
extracted from the experimental data. The dashed curve was
calculated in the additive model by Eq.~\eqref{theta:add}, and the
solid curve shows the theoretical result taking into account
additive and non-additive effects, Eq.~\eqref{theta:all}. In both
calculations the same relation between the pulse area and its power
given by Eq.~\eqref{mutau:p} was used. The solid theoretical curve
reproduces well the decrease of the phase shift from 0 to $-\pi/2$
and its saturation, confirming that indeed non-additive effects need
to be considered for a comprehensive analysis.

\subsection{Effects of high control powers}\label{sec:high}

As mentioned above, Fig.~\ref{fig:ampl} shows significant
discrepancies between experiment and theory for control powers
$P_{c}\gtrsim 5-10$ W/cm$^2$. First, Rabi oscillations are not
observed experimentally. Second, linearly and circularly polarized
control suppress the spin coherence with about same efficiency,
while the theory predicts that the suppression for circular
polarization should not exceed $50\%$.

The absence of the Rabi oscillations shows that the quantum model of
Ref.~\cite{Yugova09} used here is not fully applicable for quantum
well structures. Indeed, the classical approach developed in
Refs.~\cite{Zhukov07,astakhov08sst} shows that at high pumping
powers saturation effects become important. The classical approach
to the description of spin coherence generation and the quantum
approach of Ref.~\cite{Yugova09}, extended here to allow for
linearly polarized control pulse, coincide exactly in the limit of
weak pump and control powers~\cite{Zhukov07}. With an increase of
control power the quantum mechanical approach predicts Rabi
oscillations for the control parameters $Q_l$ and $Q_c$. The quantum
approach is justified for quantum dots where electrons and trions
preserve their coherence on the time scale of pump or control pulse.
The applicability of the quantum approach for quantum wells is
governed by the relation between the pulse duration $\tau_p$ and the
scattering time between different trion states $\tau_1$. If
$\tau_p\ll \tau_1$ the two-level model, which describes electron to
trion excitation under light pulse action, is valid. Otherwise, if
$\tau_p \gtrsim\tau_1$ the trion can scatter to another state during
the pulse action and, therefore, the Rabi oscillations become
damped.

\begin{figure}[hbt]
\includegraphics*[width=7 cm]{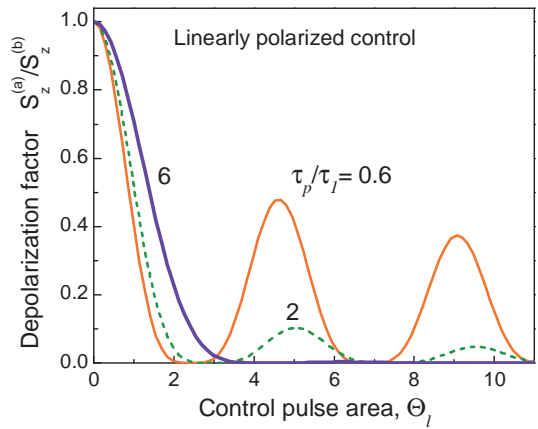}
\caption[] {(Color online) Depolarization factor for a linearly
polarized control as function of control pulse area calculated for
different ratios of pulse duration $\tau_p$ and trion scattering
time $\tau_1$: $\tau_p/\tau_1=0.6$, 2, and 6.} \label{fig:damping}
\end{figure}

To illustrate the transition from the quantum to the classical model
we performed calculations of the suppression factor
$S^{(a)}_z/S^{(b)}_z = Q_l^2$ as function of the linearly polarized
pulse area $\Theta_l$, taking into account a finite scattering time
between different trion states. We introduced it as a negative
imaginary part $-\mathrm i/(2\tau_1)$ of the trion resonance
frequency, $\omega_0$, in Eq.~\eqref{single}. The calculated
depolarization factor is shown in Fig.~\ref{fig:damping}. It is seen
that the Rabi oscillations become less pronounced with increase of
$\tau_p/\tau_1$ and eventually disappear for $\tau_p/\tau_1 \geq 3$.

The discrepancy of the theoretical predictions and the experimental
data for a circularly polarized control results from limitations of
the model. We consider the optical transition from a localized
electron state to a trion state within a two-level model neglecting
completely other excited states such as, e.g., triplet trion states,
etc.  Their inclusion, see Sec.~\ref{subsec:circ_strong} and thin solid curve in Fig.~\ref{fig:depol}, may result in the complete
suppression of the Kerr signal due to the non-additive contribution
of the circularly polarized control pulse. In addition, heating of
the electron ensemble can be considerable for pump powers exceeding
$5$~W/cm$^2$ and can cause reduction of the signal both in linear
and circular polarization.

\subsection{Efficiency of electron spin manipulation}

\begin{figure}[hbt]
\includegraphics*[width=7 cm]{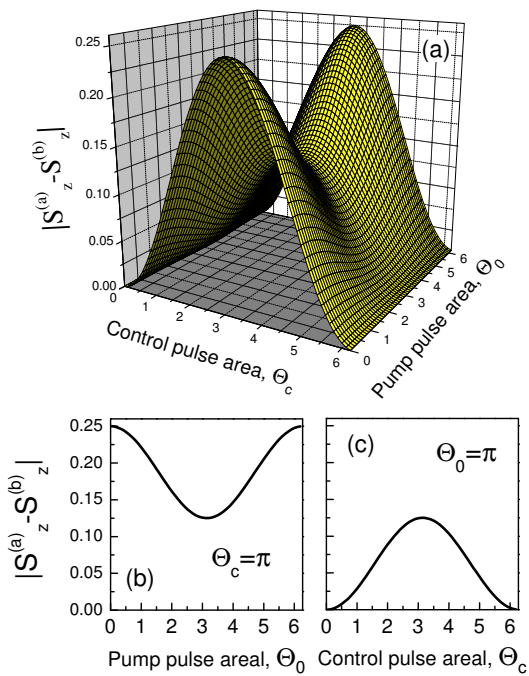}
\caption[] {(Color online) Modification of the spin $z$ component as
function of control and pump pulse areas. Panel (a) shows 3D plot.
Panels (b) and (c) show its cross-sections for $\Theta_c=\pi$ and
$\Theta_0=\pi$, respectively. Calculations performed for
co-polarized configuration and $\varphi=0$. } \label{fig:diag}
\end{figure}

It is instructive to analyze the efficiency of spin control by
circularly polarized pulses. To this end we plot in
Fig.~\ref{fig:diag} the absolute value of the spin $z$ component
change caused by the control pulse, $|S_z^{(a)} - S_z^{(b)}|$, as
function of control and pump pulse areas using
Eq.~\eqref{circ_after}. We assume that the pump and control pulses
are co-polarized and that the control pulse arrives at $\varphi=0$
of the spin beats. It is clearly seen that the modification of the
spin $z$-component is a non-monotonous function of the pump and
control pulse areas. The control efficiency depends strongly on the
pump area. For instance, if the pump area corresponds to a $\pi$
pulse, $\Theta_0=\pi$, i.e. the pump effect is maximal, the control
effect is reduced as compared with the case of $\Theta_0=0$, where
the pump is absent. This is a result of the non-additive effect of
the control pulse: if there is already some spin polarization, it is
then reduced by the non-additive effect. In other words, the
electron spin projection is limited by $|S_z|\leqslant 1/2$,
therefore, the larger is the spin created by the pump, the weaker is
the effect of the control that can be realized.

For the cross-polarized control and pump configuration (or in the
case the co-polarized control arrives at $\varphi=\pi$ of the spin
beats) the spin $z$ component modification is stronger. Indeed, the
non-additive effect of the control suppresses the spin polarization
and the spin coherence added by the control pulse has an inverse
sign as compared with the pump-induced one. Therefore, an increase
of the control pulse area from $0$ to $\pi$ always increases
$|S_z^{(a)} - S_z^{(b)}|$ independent of the pump pulse area,
contrary to the co-polarized configuration.

\section{Conclusions}

We have demonstrated experimentally the possibility to manipulate
the electron spins in quantum wells by means of polarized laser
pulses. We have shown that the coherence of resident electrons can
be increased or decreased by a circularly polarized control pulse
depending on the pump/control delay and the relative polarizations
of the pump and control pulses. This additive effect is a result of
spin coherence generation by the control pulse which may be added to
or subtracted from the pump-induced spin coherence.

Surprisingly, we have also found a non-additive effect of the
circularly polarized control pulse. This contribution is
experimentally detected by a special modulation protocol where the
control pulse is not modulated while the pump pulse is modulated and
the Kerr signal is detected by a lock-in technique.  The measured
signal is decreased by the control pulse and the suppression
efficiency is determined by the control pulse power only. It is
independent of the circular polarization of the control pulse and
the amount of spin coherence induced by the pump.

A similar suppression is observed for linearly polarized control
pulses which do not generate any spin coherence in our geometry. The
suppression efficiency is the same for linearly and circularly
polarized pulses at relatively small control powers.

The experimental findings are well explained by the proposed
theoretical model which takes into account the formation of the
singlet trion, localized on an imperfection of an $n$-type quantum
well, by polarized light. The electron spin left over from the trion
after its radiative recombination is depolarized. Since linearly
polarized light results in trion formation regardless of the
electron spin projection the spin coherence is suppressed. The model
describes both the additive and non-additive effects by circularly
polarized control pulses.

The developed model can also be applied to describe the electron
spin coherence control in quantum dots. Similarly to quantum well
systems studied here, both additive and non-additive effects of the
control pulse should be manifested in that case. One may also expect
the observation of Rabi oscillations of spin suppression for the
quantum dot systems since the trion state is much more robust and
observations of Rabi oscillations have been reported, e.g., in
experiments with optical generation of spin coherence in an ensemble
of singly charged (In,Ga)As/GaAs quantum dots~\cite{Greilich_prl}.

The manifestations of the non-additive effect are related with the
considerable spin polarization generated by the pump pulse, and in
general, do not require the trion as an intermediate state in the
spin coherence manipulation. The high spin polarization regime can
be achieved for the widely studied quantum wells containing a dense
electron gas. However, it occurs at much higher excitation densities
where other non-linear effects complicate the interpretation of the
experimental data. On the contrary, in quantum wells with a low
density electron gas as studied here a relatively high spin
polarization can be reached already at rather low excitation powers.

{\bf Acknowledgements} The authors are grateful to E.L. Ivchenko for
valuable discussions. The work was supported by the Deutsche
Forschungsgemeinschaft, the EU Seventh Framework Programme (Grant
No. 237252, Spin-optronics), the Russian Foundation for Basic
Research, the Ministry of Science and Higher Education (Poland)
through grant N202 054 32/1189 and by the Foundation for Polish
Science through subsidy 12/2007. One of the authors (MMG)
acknowledges support by the President grant for young scientists and
the ``Dynasty'' Foundation -- ICFPM.

\end{document}